\documentclass{article}
\usepackage{CJKutf8}
\usepackage[unicode]{hyperref} 
\usepackage{amsfonts}
\usepackage{amsmath}
\usepackage{amssymb}
\usepackage{amsthm}
\usepackage{youngtab}
\usepackage{braket}
\usepackage{graphicx}
\usepackage[svgnames]{xcolor}
\usepackage{ytableau,tikz}
\usepackage{here}
\usepackage{comment}
\usepackage{simplewick}
\usepackage{tikz}
\usepackage{feynmf}
\usepackage{mathtools}
\usepackage{qcircuit}

\newcommand{\ba}{\begin{eqnarray}}
\newcommand{\ea}{\end{eqnarray}}

\newcommand{\cA}{\mathcal{A}}
\newcommand{\cB}{\mathcal{B}}
\newcommand{\cC}{\mathcal{C}}
\newcommand{\cD}{\mathcal{D}}
\newcommand{\cN}{\mathcal{N}}

\newcommand{\lt}{\left}
\newcommand{\rt}{\right}

 \makeatletter
\@addtoreset{equation}{section}

\makeatother

\title{{\bf Bethe/Gauge correspodence:}\\
a short review on an aspect of the integrability nature of supersymmetric gauge theories\\
\vskip 1cm
{\bf 贝特/规范对应}\\
—超对称规范场论中可积性的一个侧面—}
\author{朱睿东 (Rui-Dong Zhu)}
\date{苏州大学高等研究院$\&$物理科学与技术学院\\
江苏省苏州市姑苏区干将路333号\\
Institute for Advanced Study \& School of Physical Science and Technology,\\ Soochow University, 333 Ganjiang Road, Gusu district, Suzhou 215006, China}

\begin{document}

\begin{CJK*}{UTF8}{gbsn}

\maketitle

\begin{center}
{\bf 摘要/abstract}
\end{center}
In this article, we provide a short review (written in Chinese) on the Bethe/Gauge correspondence. We first explain the basic idea in an explicit example of the correspondence between XXX spin chains and 2d $\cN=(2,2)$ gauge theories. The connection between 4d and 2d will then be explored by comparing the instanton and vortex partition functions. We conclude this article by briefly mentioning the similarity in the integrability structure of 4d gauge theories and 2d ones from an algebraic aspect, and a potential relation between two different integrable systems. \\

本文为一篇关于贝特/规范对应的中文短综述论文。文章将通过XXX自旋链与二维$\cN=(2,2)$规范理论之间的具体对应关系简要介绍贝特/规范对应的基本想法。随后我们将通过瞬子与涡旋配分函数之间的比较，探索二维与四维规范场论之间的关联。在文章的末尾，我们将从代数角度出发，简要提及二维与四维规范场论中可积性的相似之处，及两套可积系统之间可能存在的潜在联系。

\newpage

\noindent\hrulefill
\tableofcontents
\noindent\hrulefill



\section{序言：超对称规范场论中的可积性}

自从Seiberg与Witten于1994年发现四维$\cN=2$超对称规范场论的低能有效理论可由一条复曲线（现被称为Seiberg-Witten曲线，以下简称SW曲线）及其上定义的微分形式描述\cite{Seiberg:1994rs,Seiberg:1994aj}，可积模型的相关知识已日趋成为超对称场论研究领域的必备常识。超对称场论中最为有名并得到广泛研究的可积性源于四维$\cN=4$理论的反常维度（anomalous dimension）计算\cite{Minahan:2002ve}及其与AdS引力的对偶（参见系列综述\cite{Beisert:2010jr}）。在最先被认识到可积性的四维$\cN=2$理论中，其可积结构更为复杂因而时至今日反而得到的认知更少。SW曲线在被发现后没多久就被认识到与Hitchin系统为代表的经典可积模型对应\cite{Gorsky:1995zq,Martinec:1995by,Donagi:1995cf}，然而对于物理量计算更可控的$\Omega$-背景上的$\cN=2$理论，直到18年后的2012年才由数学家Maulik及Okounkov的工作\cite{Maulik-Okounkov}帮助我们一窥其中端倪。$\Omega$-背景上的可积模型定义在无穷维的Fock空间上，其背后的代数结构也较为复杂（参见最新的综述文献\cite{Matsuo:2023lky}），使得该领域的入门门槛较高，对初学超对称及可积性的研究者，尤其是学生极不友好。但是当我们退一步考虑$\Omega$-背景上一个参数退化的Nekrasov-Shatashivili极限时，超对称场论可被映射到最为基础的量子可积模型——自旋链模型（spin chain）\cite{Nekrasov:2009uh,Nekrasov:2009ui}。本文旨在以自旋链和超对称场论之间的对应关系——贝特/规范对应（Bethe/Gauge correspondence）为切入口对$\cN=2$理论相关的量子可积性提供一个入门级的综述。

\section{贝特/规范对应}

本文将主要关注二维规范场论与海森堡XXX自旋链之间的贝特/规范对应，并通过最简单的例子来描述这种对应关系。

考虑二维$\cN=(2,2)$超对称规范场论（由含规范场的矢量多重态与物质场对应的手性多重态构成），该类理论由规范群$G$及物质场的表示$R_j$指定。理论的低能有效物理由被称为有效扭曲超势（effective twisted superpotential）的物理量控制\cite{Nekrasov:2009uh,Nekrasov:2009ui}，其具体形式为\footnote{此处沿袭的是$Z\sim e^{-W_{\rm eff}/\epsilon},\ \epsilon\to0$的约定（其中$\epsilon$为$\Omega$背景参数或类似的红外正规化参数）。}
\begin{align}
    W_{\rm eff}(\Sigma,\{m_j\})=2\pi i\rho\cdot \sigma+\sum_j\sum_{w\in R_j}(w\cdot \sigma+m_j)\lt(\log(w\cdot \sigma+m_j)-1\rt).\label{eff-potent}
\end{align}
其中对物质场引入了被称为扭曲质量项（twisted mass）的超势，质量参数为$m_j$，$\{\sigma_i\}$为矢量多重态中（伴随表示的）标量场$\Sigma$的对角成分。在我们考虑的库伦枝\footnote{超对称规范场论的库伦枝（Coulomb branch）中，矢量多重态中的标量场在真空中获取期待值。与之相对地，物质场中的超对称标量场有非零真空期待值时，常称为希格斯枝（Higgs branch）。}中，$\sigma_i$获取非零的真空期望值，将有质量的物质场及因为$\sigma_i$的期待值而获得质量的W玻色子和它相伴的超对称费米子被积掉，便可获得上述有效扭曲超势。$\rho=\frac{1}{2}\sum_{\alpha\in\Delta_+}\alpha$为外尔向量（表示规范群中所有正根之和的一半），$w$为物质场表示中的权（weight）。当规范群中含有U(1)部分时，可以进一步在有效扭曲超势中引入Fayet-Illiopoulos项（FI项）及$\theta$-项的组合（以下简单起见，统称为FI项），
\begin{equation}
    W_{\rm FI}(\sigma)=-2\pi i\sum_j \tau_j\sigma_{{\rm U(1)}_j},
\end{equation}
其中$\tau:=\frac{\theta}{2\pi}+i\zeta$为$\theta$角与Fayet-Illiopoulos参数的组合。
例如对于$G={\rm U}(k)$，${\rm tr}\sigma=\sum_{a=1}^k\sigma_a$给出一个整体的$\mathfrak{u}(1)$因子，对应的FI项为
\begin{equation}
    W_{\rm FI}(\sigma)=-2\pi i\tau{\rm tr}\sigma.
\end{equation}
有效扭曲超势\eqref{eff-potent}可从场论的单圈计算中读取（参见例如\cite{Nekrasov:2014xaa}），类似地我们也可以从二维$\cN=(2,2)$理论的圆盘配分函数（或$\mathbb{C}$上的配分函数\footnote{这里实际上指的是$\Omega$背景上的配分函数，在特定边界条件下与圆盘配分函数相等。边界条件/边界项对于超势也有贡献，但这部分不是本文关注的重点。另外，值得注意的是当理论拥有共形对称性时，$\mathbb{C}$配分函数与球面配分函数一致，而圆盘配分函数可以视作一个半球上的配分函数。对于超势的推导，本文也将在第\ref{s:vortex}节中给予概述。}、涡旋（vortex）配分函数）获取相应信息\cite{Fujimori:2015zaa}。本文将在第\ref{s:vortex}节中简要回顾利用涡旋配分函数获得有效超势的计算方法，并借此讨论二维场论中的涡旋与四维场论中的瞬子之间的关系，从而引出第\ref{s:algebra}节中利用量子代数结构对贝特/规范对应关系一种基本物理解释。值得注意的是，对于三维$\cN=2$及四维$\cN=1$超对称理论，我们同样可以通过分别考虑$\mathbb{C}\times S^1$及$\mathbb{C}\times T^2$上的配分函数引入类似的有效超势描述理论的低能有效物理。三维及四维理论中的有效超势的数学结构也极其类似，粗略地说，我们仅需将二维有效超势中的$x\log x$部分在三维中替换为多重对数函数${\rm Li}_2(e^{-x})$，在四维中替换为椭圆版本的函数\footnote{比如对于三维理论来说，当紧化在$S^1$上时，我们需对所有的Kaluza-Klein模式求和，此时利用
\begin{equation}
    \sum_{n\in\mathbb{Z}}\lt(x+\frac{in}{R}\rt)\lt[\log\lt(x+\frac{in}{R}\rt)-1\rt]\sim\frac{1}{2\pi R}{\rm Li}_2(e^{-2\pi Rx}),
\end{equation}
可以得到高维理论中的超势\cite{Nekrasov:2009uh}。}
\begin{equation}
    g(x;p):=\sum_{k=1}^\infty \frac{e^{2\pi ik x}}{k^2(1-p^k)}-\sum_{k=1}^\infty \frac{p^ke^{-2\pi ik x}}{k^2(1-p^k)}.
\end{equation}
其中$p:=e^{2\pi i\varsigma}$，$\varsigma$为环面$T^2$上的模参数（更多细节请参考文献，例如\cite{Kimura:2020bed,Wang:2024zcr}）。

对于定义在流形$M_2$上的二维规范场论，其真空由整数的磁通量指定，对于秩为$r$的规范群，需要额外添加物理条件，
\begin{equation}
    \int_{M_2}F^a=2\pi im_a,\ m_a\in\mathbb{Z},\quad a=1,2,\dots,r.
\end{equation}
我们可以通过在作用量中增加拉格朗日乘数，
\begin{equation}
    \sum_{a=1}^rn_a\int_{M_2}F^a,\quad n_a\in\mathbb{Z},
\end{equation}
来实现该物理条件限制，它会为有效超势带来一个额外项\footnote{此处所述图像源于低能有效理论框架，但该贡献也有微观解释\cite{Nekrasov:2009uh}：由背景电场导致的粒子对生成，当它们分离至无穷远处时，会诱导一个$\theta$-项，这与此处所需的额外项一致。}：
\begin{equation}
    \Delta W_{\rm eff}(\sigma)=-2\pi i\sum_{a=1}^rn_a\sigma_a.
\end{equation}
将上述额外项的贡献纳入考虑后，二维规范场论的真空解便由下列方程给定，
\begin{equation}
    \frac{\partial W_{\rm eff}}{\partial\sigma_a}-2\pi in_a=0,\quad a=1,2,\dots,r.
\end{equation}
由于$n_a$是不受限制的任意整数，我们也可以把真空方程等价地改写为，
\begin{equation}
    \exp\lt(\frac{\partial W_{\rm eff}}{\partial\sigma_a}\rt)=1,\quad a=1,2,\dots,r.
\end{equation}
让我们来看两个后文中也会登场的简单例子。第一个例子是含有$N_f$个基本表示的物质场的U($k$)超对称规范场论，其有效扭曲超势为
\begin{equation}
    W_{\rm eff}\lt(\sigma,\{m_j\}_{j=1}^{N_f}\rt)=-2\pi i\tau\sum_{a=1}^k\sigma_a+2\pi i\sum_{a=1}^k\frac{k+1-2a}{2}\sigma_a+\sum_{j=1}^{N_f}\sum_{a=1}^k(\sigma_a+m_j)\lt(\log(\sigma_a+m_j)-1\rt),
\end{equation}
这里我们使用了外尔向量在SU($k$)群中的通用表达式，
\begin{equation}
    \rho=\lt(\frac{k-1}{2},\frac{k-3}{2},\dots,-\frac{k-3}{2},-\frac{k-1}{2}\rt).
\end{equation}
该理论的真空方程为
\begin{equation}
    (-1)^{k+1}e^{-2\pi i\tau}\prod_{j=1}^{N_f}(\sigma_a+m_j)=1,\quad a=1,2,\dots,k.\label{BAE-grass}
\end{equation}
上述方程组为一组关于$\sigma_a$之间完全独立的一元$N_f$次方程的集合，由于$\sigma_a\in\mathbb{C}$可取任意复数，对于每个真空模参数$\sigma_a$，可取$N_f$种不同的值。更进一步，（可以从涡旋配分函数的被积函数中的$\Gamma\lt((\sigma_a-\sigma_b)/\epsilon\rt)^{-1}$看到）对于$a\neq b$的任意下标，我们都有限制$\sigma_a\neq \sigma_b$，于是可得上述二维超对称规范场论的真空解个数为$\binom{N_f}{k}$。

第二个例子中，我们仍然考虑$G={\rm U}(k)$，并且在$N_f$个基本表示物质场的基础上追加1个伴随表示、$N_{\bar{f}}$个反基本表示的物质场。有效扭曲超势变为
\begin{align}
    W_{\rm eff}\lt(\sigma,m_{adj},\{m_j\}_{j=1}^{N_f},\{\bar{m}_\ell\}_{\ell=1}^{N_{\bar{f}}}\rt)=-2\pi i\tau\sum_{a=1}^k\sigma_a
    +2\pi i\sum_{a=1}^k\frac{k+1-2a}{2}\sigma_a\cr 
    +\sum_{a\neq b}(\sigma_a-\sigma_b+m_{adj})\lt(\log(\sigma_a-\sigma_b+m_{adj})-1\rt)\cr
    +\sum_{j=1}^{N_f}\sum_{a=1}^k(\sigma_a+m_j)\lt(\log(\sigma_a+m_j)-1\rt)\cr
    +\sum_{\ell=1}^{N_{\bar{f}}}\sum_{a=1}^k(-\sigma_a+\bar{m}_\ell)\lt(\log(-\sigma_a+\bar{m}_\ell)-1\rt),
\end{align}
其中我们将伴随表示物质场的质量记作$m_{adj}$，基本表示的质量记为$m_j$，反基本表示的质量写作$\bar{m}_\ell$。对应的真空方程为
\begin{equation}
    (-1)^{k+1}e^{-2\pi i\tau}\prod_{b\neq a}\frac{\sigma_a-\sigma_b+m_{adj}}{\sigma_b-\sigma_a+m_{adj}}\frac{\prod_{j=1}^{N_f}(\sigma_a+m_j)}{\prod_{\ell=1}^{N_{\bar{f}}}(-\sigma_a+\bar{m}_\ell)}=1,\quad a=1,2,\dots,k.
\end{equation}
整理得，
\begin{equation}
    e^{-2\pi i(\tau+N_{\bar{f}}/2)}\prod_{b\neq a}^k\frac{\sigma_a-\sigma_b+m_{adj}}{\sigma_a-\sigma_b-m_{adj}}\frac{\prod_{j=1}^{N_f}(\sigma_a+m_j)}{\prod_{\ell=1}^{N_{\bar{f}}}(\sigma_a-\bar{m}_\ell)}=1.\label{eq:vacua}
\end{equation}
对于可积模型熟悉的读者便会意识到上述真空方程与海森堡的XXX自旋链模型的贝特拟设方程（Bethe ansatz equation，以下简称为BAE）之间的相似性。下面为了对这个领域不太熟悉的读者，我们简要总结一下量子可积模型的构建方法及利用代数贝特拟设（algebraic Bethe ansatz）对模型求解的系统性方法。同时我们推荐想要进一步了解其中的数学及物理细节的读者参考经典的综述文献（如\cite{Faddeev:1996iy,Doikou:2009xq}等）。不过值得注意的是代数贝特拟设并非研究量子可积模型的唯一方法，坐标贝特拟设 (coordinate Bethe ansatz)、散射过程的S矩阵图像、Lax方法\cite{lax}以及非对角贝特拟设\cite{Cao:2013nza}等等，对于这些有趣角度本文只能割爱而不详述了。

量子可积模型是一类一维量子多体物理模型的统称，其主要特征是背后拥有被称为杨代数（Yangian）或量子群（quantum group）的巨大对称性代数结构，致使系统中存在的守恒量数目等于系统的自由度，从而可以帮助人们利用贝特拟设等一些精巧手段严格求解。构造量子可积模型通常由杨-Baxter方程（Yang-Baxter equation，以下简称为YBE）的一个解作为出发点。YBE的具体形式为
\begin{equation}
    {\bf R}_{12}(u-v){\bf R}_{13}(u){\bf R}_{23}(v)={\bf R}_{23}(v){\bf R}_{13}(u){\bf R}_{12}(u-v),\label{eq:YBE}
\end{equation}
其中方程的解R-矩阵，${\bf R}_{ij}(u)$，可看作两个一维准粒子$i$与$j$的两体散射过程，而YBE则告诉我们系统中的多体散射等价于多个两体散射过程。$u$通常被称为光谱参数 (spectral parameter)，是构建可积模型过程中的重要辅助参量。在散射过程的图像中，$u$也被称作速度参数 (rapidity)，可以理解为正比于散射粒子动量的物理参数，而\eqref{eq:YBE}表征的散射过程显然遵循动量守恒定律。YBE有一类非常简单的非平凡解，
\begin{equation}
    {\bf R}^{(N)}(u)=u\mathbb{I}_{2N\times 2N}+i{\cal P}^{(N)},
\end{equation}
其中$\mathbb{I}_{n\times n}$为$n\times n$尺寸的单位矩阵，${\cal P}^{(N)}$为$2N\times 2N$的矩阵，被称为置换矩阵，其功能是将两个$N$维向量空间互换，
\begin{equation}
    {\cal P}^{(N)}\vec{a}\otimes \vec{b}=\vec{b}\otimes \vec{a},\quad {\rm for}\ ^\forall\vec{a},\vec{b}\in\mathbb{C}^N.
\end{equation}
$N=2$将是本文主要关注的案例，注意到向量空间$\mathbb{C}^N$实际是每个粒子的内部自由度空间，所以$N=2$正好描述的是带有自旋$\frac{1}{2}$内部自由度的物理模型。定义$E_{ij}$为第$i$行第$j$列成分为1，其余成分为0的矩阵（即其成分为$\lt(E_{ij}\rt)_{a,b}=\delta_{i,a}\delta_{j,b}$），则${\cal P}^{(N)}$的一般表达式为
\begin{equation}
    {\cal P}^{(N)}=\sum_{i,j=1}^NE_{ij}\otimes E_{ji}.
\end{equation}
例如当$N=2$时，易得
\begin{equation}
    {\bf R}^{(2)}(u)=\lt(\begin{array}{cccc}
         u+i & 0 & 0 & 0 \\
         0 & u & i & 0 \\
         0 & i & u & 0 \\
         0 & 0 & 0 & u+i 
    \end{array}\rt).\label{R-matrix:su2}
\end{equation}
更一般地，Drinfeld和Jimbo在80年代末意识到\cite{Drinfeld:1986in,Jimbo:1985zk,Jimbo:1985vd}，每当人们构造一个被称为量子群的代数$U_q(\mathfrak{g})$，便可找到一个对应的YBE的解，即R-矩阵（\ref{s:algebra}节中将简要讲解）。这里的$q$是一个形变参数，在自旋链的背景下对应于XXX模型向XXZ模型的形变，通常也称为$q$-形变。\eqref{R-matrix:su2}中的R-矩阵对应于$\mathfrak{g}=\mathfrak{sl}_2\simeq \mathfrak{su}_2$，并且取了$q\to 1$极限的领头阶。在同样的极限下，量子群退化为杨代数。杨代数由无穷组代数元${\bf Q}^{(n)}_a$组成，其中$n\in\mathbb{N}$，而$a$则取伴随表示的下标。$\{Q_a^{(0)}\}$组成了李代数$\mathfrak{g}$，所以无论是杨代数还是量子群，均是李代数的一种自然推广。R-矩阵作为附随于量子群/杨代数出现的量，自然可以用代数元（尤其是$\mathfrak{g}$中的元）构成，对于\ref{R-matrix:su2}中的R-矩阵，我们有
\begin{equation}
    {\bf R}^{(2)}(u)=(u+i/2)\mathbb{I}_{2\times 2}\otimes \mathbb{I}_{2\times 2}+\frac{i}{2}\lt(\sigma_1\otimes \sigma_1+\sigma_2\otimes \sigma_2+\sigma_3\otimes \sigma_3\rt),
\end{equation}
其中$\sigma_{1,2,3}$为泡利矩阵。由此我们也可以明确看出R-矩阵作用在两个向量空间$\mathbb{C}^2$上，我们不妨称第一个向量空间为辅助空间（auxiliary space），而第二个空间为量子空间（quantum space），则R-矩阵也可作为辅助空间上的$2\times 2$矩阵，用量子空间中的算符表达：
\begin{equation}
    {\bf R}^{(2)}(u)=\lt(\begin{array}{cc}
       u+\frac{i}{2}+iJ_z  & iJ_- \\
       iJ^+  & u+\frac{i}{2}-iJ_z
    \end{array}\rt),\label{R-operator}
\end{equation}
此处$J_z:=\frac{\sigma_3}{2}$, $J_\pm:=\frac{\sigma_1\pm i\sigma_2}{2}$为$\mathfrak{g}=\mathfrak{su}_2$的生成元，
\begin{equation}
    [J_z,J_\pm]=\pm J_\pm,\quad [J^+,J^-]=2J_z.
\end{equation}
到了这里，我们就可以容易地将量子空间从$\mathbb{C}^2$替换为$\mathbb{C}^{2s}$（$s$为半整数），即对应地考虑$J_z,J_\pm$的自旋$s$表示，这样更为一般的算符称之为Lax算符，记为${\bf L}(u)$。与YBE平行地，Lax算符满足以下的RLL关系：
\begin{equation}
    {\bf R}^{(2)}_{00'}(u-v){\bf L}_{0i}(u){\bf L}_{0'i}(v)={\bf L}_{0'i}(v){\bf L}_{0i}(u){\bf R}^{(2)}_{00'}(u-v),\label{eq:RLL}
\end{equation}
其中$0$和$0'$分别表示两个辅助空间，$i$表示格点模型中的某个量子空间（或第$i$个粒子的内部空间）。对于长度为$L$的一维格点上的量子可积模型，我们在每个格点上取一个量子空间，于是可以定义以下被称为monodromy矩阵的量：
\begin{equation}
    T(u):={\bf L}_{0L}(u){\bf L}_{0(L-1)}(u)\dots {\bf L}_{01}(u).\label{def-mono}
\end{equation}
当我们对格点模型施加周期性边界条件时，可以通过对monodromy矩阵的辅助空间取迹来构建极为重要的物理量——转移矩阵（transfer matrix），
\begin{equation}
    t(u):={\rm tr}_0T(u).
\end{equation}
利用RLL关系\eqref{eq:RLL}可以巧妙地证明不同参数的转移矩阵之间都互相对易，
\begin{equation}
    [t(u),t(v)]=0,\quad ^\forall u,v.
\end{equation}
因此，当我们将转移矩阵$t(u)$关于$u$或$u^{-1}$展开，得到的算符值系数全部对易，比如设
\begin{equation}
    t(u)=t(0)\exp\lt(\sum_{i=1}^\infty H^{(i)}u^i\rt),
\end{equation}
则
\begin{equation}
    [H^{(i)},H^{(j)}]=0,\quad ^\forall i,j.
\end{equation}
其中$H^{(1)}$特别有趣，当每个格点上的自旋为$s=\frac{1}{2}$时，我们可以直接使用R-矩阵作为Lax算符计算得，
\begin{align}
    H^{(1)}=\lt.\frac{{\rm d}}{{\rm d}u}\log t(u)\rt|_{u=0}=\frac{1}{2}\sum_{j=1}^L\lt(\sigma_1^{(j)}\otimes \sigma_1^{(j+1)}+\sigma_2^{(j)}\otimes \sigma_2^{(j+1)}+\sigma_3^{(j)}\otimes \sigma_3^{(j+1)}\rt)+\frac{L}{2},\label{XXX-Ham}
\end{align}
在上式中，$\sigma_k^{(j)}$代表了第$j$个量子空间上定义的泡利矩阵，并且由于转移矩阵的定义中的迹${\rm tr}_0$施加了周期性边界条件，第$L+1$个量子空间被等同视为格点中的第$1$个量子空间。除去\eqref{XXX-Ham}中最后的常数项，$H^{(1)}$与海森堡在1928年提出的一维磁性模型\cite{Heisenberg}完全一致，而正是在海森堡模型中，1931年贝特使用了他关于波函数的拟设解法，完整求解了模型的能谱\cite{Bethe:1931hc}。这种方法现在被称为贝特拟设（Bethe ansatz），并在70年代末80年代初由列宁格勒学派提炼为代数贝特拟设方法（\cite{Faddeev:1996iy}提供了最标准的综述文献），可以对几乎所有周期性边界条件的量子可积模型适用。在简要阐述代数贝特拟设之前，让我们先对海森堡模型与\eqref{R-matrix:su2}中的R-矩阵给出的量子可积模型之间的关系作一个补充说明。贝特拟设需要从所有自旋均朝上的态（容易验证其一定为$H^{(1)}$的本征态）
\begin{equation}
    \ket{\Omega}=\bigotimes_{j=1}^L\ket{\uparrow},
\end{equation}
出发，系统性地寻找所有的本征态。对于海森堡磁性模型${\cal H}=JH^{(1)}$, $J$为耦合常数。当$J<0$，即铁磁模型时，$\ket{\Omega}$为系统的基态，而对于$J>0$的反铁磁性模型时，$\ket{\Omega}$为能量最高的态，系统基态附近的物理信息需要更为详尽的分析才能获取。反铁磁海森堡模型中最有趣的现象便是Haldane猜想\cite{Haldane:1982rj,Haldane:1983ru}，其中最重要的一点便是在热力学极限$L\to\infty$下，反铁磁海森堡模型的基态与第一激发态之间是否存在能隙与模型中自旋为整数还是半整数密切相关。自旋为半整数时，反铁磁模型并不存在能隙，自旋为整数时，则有非零能隙。自旋为$1/2$时，可以用贝特拟设方法验证Haldane的猜想，对于自旋取其它值时，我们仍可用Lax算符的不同表示来构建高自旋的可积模型，但是比如$s=1$时，所得的哈密顿量与海森堡模型并不完全一致\cite{Babujian:1983ae}：
\begin{equation}
    H^{(1)}=\frac{1}{4}\sum_{j=1}^{L}\left(\vec{S}^{(j)}_1\cdot \vec{S}^{(j+1)}_{1}+\beta(\vec{S}^{(j)}_1\cdot \vec{S}^{(j+1)}_{1})^2\right),\quad \beta=1,\label{int-s-1}
\end{equation}
这里$\vec{S}_1$为$\mathfrak{su}_2$的生成元在$s=1$时的表示。\eqref{int-s-1}所给出的模型是无能隙的，但是当其中的$\beta=\frac{1}{3}$时，这对应另一个可严格求解的模型——Affleck-Kennedy-Lieb-Tasaki模型（简称AKLT模型），并且该模型具有非零能隙\cite{Affleck:1987vf}。当前人们的普遍理解为，在$\frac{1}{3}\leq \beta<1$的某处发生了相变，$\beta=0$的海森堡模型以及$\beta=\frac{1}{3}$的AKLT模型都处在一个被称为Haldane phase的拓扑相之中，在该物相中，一种非局域的隐藏序参量所对应的对称性发生破缺，这与当今备受瞩目的对称性保护拓扑相（symmetry protected topological phase）及广义对称性（generalized symmetry）都有密不可分的联系。

回到代数贝特拟设方法，我们仍然将焦点放在自旋$s=\frac{1}{2}$的情况上。先将monodromy矩阵在辅助空间上写作
\begin{equation}
    T(u)=\lt(\begin{array}{cc}
       \cA(u)  & \cB(u)  \\
        \cC(u) & \cD(u)
    \end{array}\rt),
\end{equation}
这里$\cA,\cB,\cC,\cD$均为作用在$L$个量子空间上的巨大矩阵算符。可以通过迭代计算发现$\cA$与$\cD$并不改变量子态的总自旋，而$\cB$和$\cC$分别使总自旋下降或升高$1$，可以被认为是升/降算符。它们之间的对易关系可由RLL关系\eqref{eq:RLL}算出，对于任意谱参数$^\forall u_1,u_2$，$[\cB(u_1),\cB(u_2)]=0$，$[\cC(u_1),\cC(u_2)]=0$，
\begin{equation}
    \begin{aligned}
        &(u_1-u_2+i){\cal B}(u_1){\cal A}(u_2)=(u_1-u_2){\cal A}(u_2){\cal B}(u_1)+i{\cal B}(u_2){\cal A}(u_1)\,,
\\
    &(u_1-u_2+i){\cal B}(u_2){\cal D}(u_1)=(u_1-u_2){\cal D}(u_1){\cal B}(u_2)+i{\cal B}(u_1){\cal D}(u_2)\,,
\\
    &(u_1-u_2+i){\cal C}(u_2){\cal A}(u_1)=(u_1-u_2){\cal A}(u_1){\cal C}(u_2)+i{\cal C}(u_1){\cal A}(u_2)\,,
\\
    &(u_1-u_2+i){\cal C}(u_1){\cal D}(u_2)=(u_1-u_2){\cal D}(u_2){\cal C}(u_1)+i{\cal C}(u_2){\cal D}(u_1)\,,\label{ABCD}
    \end{aligned}
\end{equation}
并且可以很容易从\eqref{R-operator}中通过归纳法得出$\cA$和$\cD$算符在$\ket{\Omega}$上的作用为：
\begin{align}
    {\cal A}(u)\ket{\Omega}=\prod_{l=1}^L\lt(\lt(u+\frac{i}{2}\rt)+iJ_{l}^z\rt)\ket{\Omega},\\
    {\cal D}(u)\ket{\Omega}=\prod_{l=1}^L\lt(\lt(u+\frac{i}{2}\rt)-iJ_{l}^z\rt)\ket{\Omega}.
\end{align}
如果能将转移矩阵$t(u)=\cA(u)+\cD(u)$完全对角化并计算出它的本征值，则包括哈密顿量在内的一大类守恒量的本征值都能得到求解。代数贝特拟设唯一的假设便是转移矩阵的本征态都由升算符$\cB$作用在$\ket{\Omega}$上生成，
\begin{equation}
    \ket{\psi(\{u_k\}_{k=1}^M)}=\prod_{k=1}^M\cB(u_k)\ket{\Omega},\label{ansatz}
\end{equation}
这里每个升算符$\cB$都会将态的总自旋降低1，而这个过程其实等效地在自旋链上激发了一个准粒子，通常称为磁子（magnon），上式中的$M$代表磁子的个数。
我们可以利用对易关系\eqref{ABCD}计算转移矩阵在上述态上的作用，
\begin{equation}
    t(u)\ket{\psi(\{u_k\}_{k=1}^M)}=\lt[(u+i)^L\prod_{k=1}^M\frac{u_k-u+i}{u_k-u}+u^L\prod_{k=1}^M\frac{u-u_k+i}{u-u_k}\rt]\ket{\psi(\{u_k\}_{k=1}^M)}+\dots\label{t-action}
\end{equation}
$\dots$表示对于任意的态$\ket{\psi(\{u_k\}_{k=1}^M)}$，转移矩阵的作用会有非对角成分，因而我们需要对\eqref{ansatz}中的$u_k$加一些限制条件使得多余的非对角项消失，这种限制便是贝特拟设方程BAE。这边我们走一条推导BAE的捷径：设\eqref{t-action}中的多余项为0，由于$t(u)$的表达式是$u$的多项式，我们自然期待$t(u)$的本征值也是$u$的多项式，但\eqref{t-action}的表达式显然有许多位于$u=u_k$的极点，所以我们需要要求这些极点处的留数为0，即
\begin{equation}
    \frac{(u_j+i)^L}{u_j^L}\prod_{k\neq j}\frac{u_j-u_k-i}{u_j-u_k+i}=1.\label{eq:BAE}
\end{equation}
这就是著名的BAE，其中我们假设了所有的$u_k$均取不同的值，而由于类似泡利不相容原理的表示论原因，在自旋$s=\frac{1}{2}$的模型中作用两次相同的升算符$\cB(u_k)$，波函数的确会自动变为0。感兴趣的读者也可以进一步利用对易关系\eqref{ABCD}计算\eqref{t-action}中的多余项，并检验当BAE成立时，多余项的确会自动消失。

至此，我们可以看到自旋链模型中的贝特拟设方程\eqref{eq:BAE}与超对称规范场论中的真空方程\eqref{eq:vacua}形式非常类似。其实还可以更进一步考虑更为一般的转移矩阵，
\begin{align}
    t(u,\theta;\{s_l\},\{\vartheta_l\}):={\rm tr}_0\lt({\rm Ph}_0(\theta){\bf L}_{0L}(u-\vartheta_L){\bf L}_{0(L-1)}(u-\vartheta_{L-1})\dots {\bf L}_{01}(u-\vartheta_1)\rt),
\end{align}
其中${\rm Ph}(\theta)={\rm diag}(e^{i\theta/2},e^{-i\theta/2})$，并且我们在每个量子空间上分别取自旋为$s_i$的表示。$\theta$被称为扭角（twisted angle），在对应的自旋链引入一个扭曲周期性边界条件：
\begin{equation}
    J_z^{(L+1)}=J_z^{1},\quad J^{(L+1)}_{\pm}=e^{\pm i\theta}J^{(1)}_{\pm}.
\end{equation}
不难确认，新模型的BAE为：
\begin{equation}
    e^{i\theta}\prod_{l=1}^L\frac{u_j+i/2+is_l-\vartheta_l}{u_j+i/2-is_l-\vartheta_l}\prod_{k\neq j}^M\frac{u_j-u_k-i}{u_j-u_k+i}=1.\label{eq:t-BAE}
\end{equation}
当规范场论中$N_f=N_{\bar{f}}$时，我们便可建立规范场论与自旋链模型之间的对应词典：
\begin{align}
    \begin{tabular}{|c|c|}
    \hline
      自旋链   & 超对称规范场论  \\
      \hline
       $L$  & $N_f=N_{\bar{f}}$\\
       \hline
       $M$ & $k$\\
       \hline
       $u_a/i$ & $\sigma_a/m_{adj}$\\
       \hline
       $\theta$ & $2\pi (\tau+N_{\bar{f}}/2)$ \\
       \hline
       $1/2-s_l-\vartheta_l/i$ & $m_l/m_{adj}$\\
       \hline
       $1/2+s_l-\vartheta_l/i$ & $\bar{m}_l/m_{adj}$\\
       \hline
    \end{tabular}\label{dictionary}
\end{align}
这里就是Nekrasov-Shatashvili在\cite{Nekrasov:2009uh,Nekrasov:2009ui}中提出的所谓贝特/规范对应关系（Bethe/Gauge correspondence）的基础。

\subsection*{自旋链中物理量在规范场论中的实现}

表\eqref{dictionary}中建立的贝特/规范对应关系仅仅是确立两种看似毫无关联的模型之间联系的最基础的一步，实际这种对应关系更为深刻。目前为止的文献中，主要关注的是如何在超对称规范场论中实现自旋链模型的各种物理量，这里由于篇幅限制，我们仅列举文献中已知的一些结果（更加详细的综述会在后续加长版本的文章中呈现）。比如自旋链中含$k$个磁子的物理态（即BAE的物理解）个数与超对称规范场论中计数真空态个数的Witten指数匹配\cite{Shu:2022vpk,Shu:2024crv}，本征态的波函数与带orbifold defect的配分函数对应\cite{Bullimore:2017lwu}，而自旋链模型中的Baxter Q-函数$Q(u):=\prod_{j=1}^M(u-u_j)$、哈密顿量、R-矩阵等重要物理量都可以从规范场论对应的Nakajima quiver variety及相关联的（带缺陷）涡旋配分函数中构造\cite{Pushkar:2016qvw,Aganagic:2017gsx}（参见类似的物理角度的研究工作\cite{Gu:2022dac,Gu:2022ugf}等）。另外\cite{Gu:2022ugf}中的一个非常有趣的视角是，二维超对称理论中的畴壁（domain wall）之间的散射S矩阵呈现的可积性（参考早期工作\cite{Fendley:1990zj,Fendley:1992dm}）与YBE的散射过程图像也恰好吻合。

\section{从四维到二维：瞬子与涡旋}\label{s:vortex}

前文在计算二维超对称规范场论的有效扭曲超势及自旋链中的许多物理量时都提到了涡旋配分函数，事实上涡旋模空间上的量子几何（quantum geometry）结构可以用于系统性地构建量子可积模型。而更为有趣的是，二维规范场论中的涡旋可以被视作四维$\cN=2$超对称规范场论中瞬子（instanton）的一个特殊极限\cite{Hanany:2003hp,Eto:2005yh,Eto:2006pg,Bonelli:2011fq,Dorey:2011pa,Chen:2011sj,Fujimori:2015zaa}，也就是说二维超对称理论里看到的量子可积性可以被理解为四维超对称理论里更为宏大的可积性图像中的一个特殊情况\footnote{此处仅为作者的个人观点，也有学者认为可积性来源于二维的特殊性，而四维理论的可积性与它们可以约化至二维有关。但是需要注意的是，这里所说的瞬子模空间中的可积性与前面提到的将$x\log x$函数替换为其椭圆版的（与XYZ自旋链对偶的）可积性看似并无关联。}。这一节中，我们尝试简单地描述二维涡旋配分函数与四维瞬子配分函数之间的关系。

和更为有名的瞬子配分函数的计算类似，二维$\cN=(2,2)$超对称规范场论在复平面$\mathbb{C}$上的配分函数会因为涡旋的贡献而发散，因此我们需要引入$\Omega$背景作为红外正规化方案。我们可以等效地考虑以下四维空间$\mathbb{C}_\epsilon\times T^2$上的$\cN=1$超对称理论，
\begin{equation}
    {\rm d}s^2=\lt|{\rm d}z-iz(\epsilon {\rm d}w+\bar{\epsilon}{\rm d}\bar{w})\rt|^2+|{\rm d}w|^2,
\end{equation}
其中$z$为二维空间$\mathbb{C}$上的复坐标，$w$为$T^2$上的坐标（$w\sim w+2\pi (R_3+iR_4)$）。将四维理论约化至二维后，便得到了定义在二维$\Omega$背景上的$\cN=(2,2)$超对称规范场论，当然在$\epsilon\to 0$的极限下，$\Omega$背景会变为普通的平坦二维时空。利用局域化（localization）的计算手法，得到规范群为$G$，含$R_j$表示的物质场的二维理论的涡旋配分函数的具体表达式为\cite{Fujimori:2015zaa}：
\begin{equation}
    Z_{\vec{a},\vec{\mu}}=\oint_{C_{\vec{a},\vec{\mu}}}\lt(\prod_{i=1}^r\frac{{\rm d}\sigma_i}{2\pi i\epsilon}\rt)\exp\lt(-\frac{2\pi i\sigma\cdot\tau}{\epsilon}\rt)\prod_{\alpha\in \Delta}\Gamma\lt(\frac{\alpha \cdot \sigma}{\epsilon}\rt)^{-1}\prod_{j}\prod_{w\in R_j}\Gamma\lt(\frac{w\cdot\sigma+m_j}{\epsilon}\rt),
\end{equation}
这里$\vec{a}=(a_1,a_2,\dots,a_r)$ （$r$为规范群的秩）为所有物质场中挑选出的$r$个标签，与之对应地，我们在$R_{a_i}$表示中分别选取一个权$\mu_i$。圈积分的路径包含所有$\mu_i\cdot\sigma=-m_{a_i}-\epsilon k_i$的极点，其中$k_i\in\mathbb{N}$为任意自然数，代表了涡旋的卷绕数（winding number）。该配分函数可以写成关于$\vec{k}=(k_1,k_2,\dots,k_r)$的留数和，$Z_{\vec{a},\vec{\mu}}=\sum_{\vec{k}}Z_{\vec{a},\vec{\mu},\vec{k}}$，当规范群的秩较低时，我们可以进一步给出涡旋配分函数的闭合表达式。比如带$N$个物质场的U(1)理论，其配分函数为
\begin{align}
    Z_a&=\oint_{C_a}\frac{{\rm d}\sigma}{2\pi i\epsilon}e^{-2\pi i\sigma\tau/\epsilon}\prod_{j=1}^N\Gamma\lt(\frac{\sigma+m_j}{\epsilon}\rt)\cr
    &=e^{2\pi i\tau m_a/\epsilon}\prod_{j\neq a}\Gamma\lt(\frac{m_j-m_a}{\epsilon}\rt)\sum_{k=0}^\infty e^{2\pi i\tau k}\prod_{j}\prod_{l=1}^k\frac{\epsilon}{m_j-m_a-l\epsilon}\cr
    &=e^{2\pi i\tau m_a/\epsilon}\prod_{j\neq a}\Gamma\lt(\frac{m_j-m_a}{\epsilon}\rt)\ _0F_{N-1}\lt(\{1+(m_a-m_j)/\epsilon\}_{j=1,j\neq a}^{N};(-1)^Ne^{2\pi i\tau}\rt),\label{vortex-fund}
\end{align}
这里我们取了$\sigma=-m_a-k\epsilon$的极点对应的留数。我们注意到配分函数的最终表达式——广义超几何函数满足微分方程，
\begin{equation}
    \lt[z\frac{{\rm d}}{{\rm d}z}\prod_{j=1}^q\lt(z\frac{{\rm d}}{{\rm d}z}+b_j-1\rt)-z\rt]\ _0F_{q}(\{b_j\}_{j=1}^q;z)=0,
\end{equation}
因此U(1)理论的涡旋配分函数满足微分方程\footnote{注意$Z_a$中的整体系数$e^{2\pi i\tau m_a/\epsilon}$造成的微分项的平移。}，
\begin{equation}
    \lt[\prod_{j=1}^N\lt(-\frac{\epsilon}{2\pi i}\frac{{\rm d}}{{\rm d}\tau}+m_j\rt)-\epsilon^Ne^{2\pi i\tau}\rt]Z_a=0.\label{ODE-U1}
\end{equation}
这个方程其实可以通过挪动积分变量$\sigma\to\sigma+\epsilon$推导：由于移动过程中圈积分路径$C_a$不会穿过任何被积函数的极点，所以积分结果不变，但是其积分表达式会变为
\begin{equation}
    Z_a=e^{-2\pi i\tau}\oint_{C_a}\frac{{\rm d}\sigma}{2\pi i\epsilon}e^{-2\pi i\sigma\tau/\epsilon}\prod_{j=1}^N\frac{\sigma+m_j}{\epsilon}\Gamma\lt(\frac{\sigma+m_j}{\epsilon}\rt).
\end{equation}
显然由于对$Z_a$作用$\hat{\sigma}:=-\frac{\epsilon}{2\pi i}\frac{{\rm d}}{{\rm d}\tau}$会在被积函数中生成一个$\sigma$，从上面的等价积分表达式可以容易地看出\eqref{ODE-U1}成立。这种推导方法对非阿贝尔规范群同样有效，但我们可以选择对哪些$\sigma_i$的组合进行正整数倍的$\epsilon$的平移来推导对应的微分方程。例如，对于带$N$个基础表示物质场的U($k$)理论，我们考虑对其中单个$\sigma_j$平移1个$\epsilon$，对应的微分方程是
\begin{equation}
    \lt[\prod_{m\neq j}^k(\hat{\sigma}_m-\hat{\sigma}_j)\prod_{l=1}^N(\hat{\sigma}_j+m_l)-(-1)^{k-1}\epsilon^{N}e^{2\pi i\tau_j}\prod_{m\neq j}^k(\hat{\sigma}_m-\hat{\sigma}_j-\epsilon)\rt]Z_{\vec{a},\vec{\mu}}=0.\label{quant-BAE-grass}
\end{equation}
这个方程在$\epsilon\to 0$，并且将$\hat{\sigma}$替换成普通参数$\sigma$的极限下与真空方程\eqref{BAE-grass}（在对FI参数$e^{2\pi i\tau_j}$进行重整化吸收$\epsilon^N$后）一致。换句话说，这里推导的微分方程可以看作对\eqref{BAE-grass}给出的几何曲线的量子化，也就是一种量子几何。同样，对于含有一个伴随表示及$N$个基本表示、$N$个反基本表示物质场的U($k$)理论，我们得到BAE\eqref{eq:vacua}的量子化版本，
\begin{align}
    &\lt[\prod_{m\neq j}^k(\hat{\sigma}_j-\hat{\sigma}_m+m_{adj})(\hat{\sigma}_m-\hat{\sigma}_j)\prod_{l=1}^N(\hat{\sigma}_j+m_l)\rt.\cr
    &\lt.-(-1)^{N}\epsilon^{N}e^{2\pi i\tau_j}\prod_{m\neq j}^k(\hat{\sigma}_j-\hat{\sigma}_m-m_{adj}+\epsilon)(\hat{\sigma}_m-\hat{\sigma}_j-\epsilon)\prod_{l=1}^N(\hat{\sigma}_j-\bar{m}_l+\epsilon)\rt]Z_{\vec{a},\vec{\mu}}=0.
\end{align}

极其类似地，四维$\cN=2$超对称规范场论的低能有效物理由被称为Seiberg-Witten曲线（SW曲线）的复曲线及对应的前势（prepotential）决定。U($N$)理论的SW曲线形式为：
\begin{equation}
    P_a(z)y+\mathfrak{q}\frac{P_f(z)}{y}=P_N(z),
\end{equation}
此处$P_f$, $P_a$分别为含有基础表示和反基础表示物质场信息的多项式，$P_N$则是（间接）含有库伦枝参数等矢量多重态信息的$N$次多项式。我们可以调节理论中的库伦枝参数到特殊值从而还原出二维理论的真空方程。比如我们考虑四维U($N$)理论带$N$个基础表示，其中对参数调节实现（更为详细的讨论参见\cite{Dorey:2011pa}），
\begin{equation}
    P_N(z)=-\mathfrak{h}+(\mathfrak{h}+2)P_f(z),\quad \mathfrak{q}=-\mathfrak{h}(\mathfrak{h}+2),
\end{equation}
在此情况下，SW曲线退化为
\begin{equation}
    \lt[y-(\mathfrak{h}+2)P_f(z)\rt](y+\mathfrak{h})=0.
\end{equation}
由于$P_f(z)=\prod_{j=1}^N(z-m_j)$，上述曲线的一部分便可改写为仅含$N$个物质场的二维理论的真空方程\eqref{BAE-grass}，其中$(\mathfrak{h}+2)y^{-1}\equiv (-1)^{k+1}e^{-2\pi i\tau}$。同样的讨论可以被推广到含有反基本表示和伴随表示物质场的理论中。更进一步地，我们也可以对SW曲线（对应的经典可积系统）进行量子化\cite{Nekrasov:2009rc}\footnote{数学上存在几何量子化（geometric quantization）与形变量子化（deformation quantization）之分，前者与量子力学中的量子化一致，后者其实是复流形的全纯函数环上的形变。这里所说的量子化更类似于形变量子化（参考\cite{Gukov:2008ve,Gukov:2010sw}），不过从四维/五维或者拓扑弦的角度这也比较自然，因为SW曲线源于弦论构造中涉及的Calabi-Yau流形的镜像曲线（mirror curve）。}，其可以在（将四维$\Omega$背景上的一个参数$\epsilon_2$取为0的）Nekrasov-Shatashvili极限下实现，而量子化方案由余下的有限$\Omega$背景参数$\epsilon_1=\hbar$决定：$\lt[\hat{z},\log y\rt]=\hbar$。这和前面提到的二维$\Omega$背景上的真空方程量子化方案几乎如出一辙：$\hat{z}\sim \hat{\sigma}=-\frac{\epsilon}{2\pi i}\frac{{\rm d}}{{\rm d}\tau}\sim \hbar\frac{{\rm d}}{{\rm d}\log y}$，而且两个理论中的$\Omega$背景参数也相互匹配。当然我们得注意到二维理论中量子化版本的方程，例如\eqref{quant-BAE-grass}中蕴含了非阿贝尔规范场的贡献，这是无法直接从SW曲线中读取出的信息。

更具体地，我们可以直接计算证明，四维$\cN=2$理论的瞬子配分函数在调整库伦枝参数的特殊极限下与二维理论的涡旋配分函数一致。让我们先来看微扰部分，简单起见，我们仍然从四维带$N$个基本表示物质场的U($N$)理论出发，其经典与单圈微扰部分为
\begin{equation}
    Z_{\rm pert.}=\exp\lt(-\frac{\pi i\tau_0}{\epsilon_1\epsilon_2}\vec{a}\cdot \vec{a}\rt)\prod_{j\neq m}\Gamma_2(a_j-a_m|\epsilon_1,\epsilon_2)^{-1}\prod_{j,l}^N\Gamma_2(a_j-m_l|\epsilon_1,\epsilon_2),
\end{equation}
其中$\Gamma_2(x|\epsilon_1,\epsilon_2):=\prod_{n,m=0}^\infty\frac{1}{x+n\epsilon_1+m\epsilon_2}$，并且满足
\begin{equation}
    \frac{\Gamma_2(x|\epsilon_1,\epsilon_2)}{\Gamma_2(x+\epsilon_1|\epsilon_1,\epsilon_2)}=\frac{\epsilon_2^{\frac{x}{\epsilon_2}-\frac{1}{2}}}{\sqrt{2\pi}}\Gamma\lt(\frac{x}{\epsilon_2}\rt).
\end{equation}
当我们将库伦枝参数取$a_j\to -m_j-\epsilon_1\delta_{j,a}$的极限并计算$Z_{\rm 1-loop}$的留数时，可以（去除一些不重要的整体系数）重现出二维U(1)理论带$N$个基本表示物质场时的微扰贡献，
\begin{equation}
    Z_{\rm pert.}\to \prod_{l\neq a}\Gamma\lt(\frac{m_l-m_a}{\epsilon_2}\rt),
\end{equation}
其中四维$\Omega$背景上的一个参数$\epsilon_2$转化为了二维$\Omega$背景参数。在同样的极限下，四维配分函数的瞬子修正部分退化为二维配分函数的涡旋贡献。瞬子配分函数的积分表达式为
\begin{align}
    Z_{\rm instanton}=\sum_{n=1}^\infty\mathfrak{q}^n\lt(\frac{\epsilon_1+\epsilon_2}{\epsilon_1\epsilon_2}\rt)^n\oint_{C_{JK}}\prod_{\alpha=1}^n\frac{{\rm d}\Phi_\alpha}{2\pi i}\prod_{j=1}^N\frac{(\Phi_\alpha +m_j)}{(-\Phi_\alpha+a_j)(\Phi_\alpha-a_j+\epsilon_1+\epsilon_2)}\cr
    \times \prod_{\beta<\alpha}\frac{(\Phi_\alpha-\Phi_\beta)^2\lt[(\Phi_\alpha-\Phi_\beta)^2-(\epsilon_1+\epsilon_2)^2\rt]}{\lt[(\Phi_\alpha-\Phi_\beta)^2-\epsilon_1^2\rt]\lt[(\Phi_\alpha-\Phi_\beta)^2-\epsilon_2^2\rt]},
\end{align}
这里的积分路径$C_{JK}$遵循JK处方（Jeffery-Kirwan prescription），当规范群为U($N$)时，积分路径选取由$N$个杨图中的箱子组成的极点集合：
\begin{equation}
    \lt\{\phi_{b,(i,j)}=a_b+(i-1)\epsilon_1+(j-1)\epsilon_2\rt\},
\end{equation}
其中$b$是杨图的编号，$(i,j)$则对应该杨图中第$i$行第$j$列的一个箱子。然而当配分函数中取$a_j\to -m_j-\epsilon_1\delta_{j,a}$的极限时，JK处方中极点$\phi_{b\neq a,(1,1)}=a_b\to -m_b$。由于$(\Phi_\alpha+m_b)$项的存在，留数将变为$0$，同样$\phi_{a,(2,1)}=a_a+\epsilon_1$处的留数也会在此极限下消失。这意味着除了第$a$个杨图可以有一行，其余杨图都必须为空，也就是说我们仅取$\phi_j=-m_a-\epsilon_1+(j-1)\epsilon_2$（$j=1,2,\dots,n$）处的留数。因此，在$a_j\to -m_j-\epsilon_1\delta_{j,a}$的极限下瞬子配分函数简化为
\begin{align}
    Z_{\rm instanton}\to \sum_{n=1}^\infty(-1)^{n-1}\mathfrak{q}^n\prod_{j=1}^n\frac{1}{j\epsilon_2}\prod_{b\neq a}\frac{1}{m_b-m_a-j\epsilon_2}.
\end{align}
与\eqref{vortex-fund}式比较便会发现，瞬子配分函数与涡旋配分函数在$\epsilon\equiv \epsilon_2$，$\mathfrak{q}/\epsilon_2\equiv \epsilon e^{2\pi i\tau}$的词典下一致。类似地，我们可以考虑极限$\vec{a}\to -\vec{m}-\vec{n}\epsilon_1$并在瞬子配分函数中加上反基本表示的贡献，从而得到含（质量为$\epsilon_1$的）伴随表示、基本和反基本表示物质场二维理论的涡旋配分函数。由于篇幅限制，请感兴趣的读者参考\cite{Fujimori:2015zaa}及相关文献。

\section{模空间上的量子可积性与代数结构}\label{s:algebra}

上一节中，我们从配分函数的角度阐述了四维超对称规范场论中的瞬子与二维超对称理论中的涡旋之间的关联，这也可以从ADHM（Atiyah-Drinfeld-Hitchin-Manin）型的瞬子解及涡旋解之间的关系看出。本节中，我们进一步简要说明在ADHM解的模空间上如何得到量子可积模型的数据，以及这背后有趣而意义深远的量子代数结构。

对于四维理论中的瞬子解，ADHM四人给出的著名构建方法基于以下数据\cite{Atiyah:1978ri}：两个向量空间$\mathbb{C}^n$、$\mathbb{C}^N$，以及向量空间之间的映射$B_1,B_2\in{\rm Hom}(\mathbb{C}^n,\mathbb{C}^n)$、$I\in {\rm Hom}(\mathbb{C}^N,\mathbb{C}^n)$、$J\in {\rm Hom}(\mathbb{C}^n,\mathbb{C}^N)$构成。这些数据可以画成以下箭图（quiver），
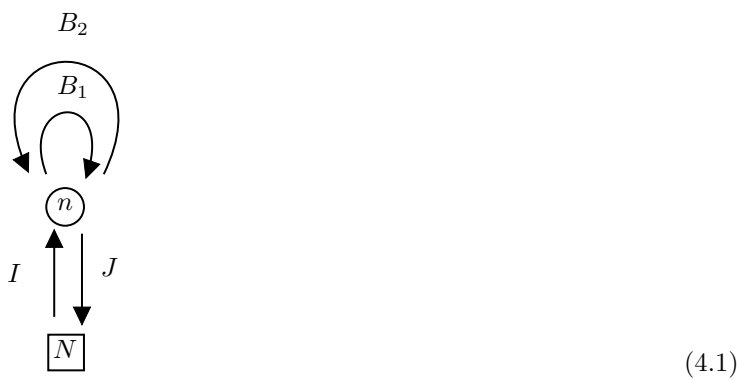
\begin{align}
\tikzset{every picture/.style={line width=0.75pt}} 
\begin{tikzpicture}[x=0.75pt,y=0.75pt,yscale=-1,xscale=1]
\draw   (286,150.5) .. controls (286,145.25) and (290.25,141) .. (295.5,141) .. controls (300.75,141) and (305,145.25) .. (305,150.5) .. controls (305,155.75) and (300.75,160) .. (295.5,160) .. controls (290.25,160) and (286,155.75) .. (286,150.5) -- cycle ;
\draw   (287,215) -- (305,215) -- (305,233) -- (287,233) -- cycle ;
\draw    (290,206) -- (290,165) ;
\draw [shift={(290,162)}, rotate = 90] [fill={rgb, 255:red, 0; green, 0; blue, 0 }  ][line width=0.08]  [draw opacity=0] (8.93,-4.29) -- (0,0) -- (8.93,4.29) -- cycle    ;
\draw    (304,164) -- (304,207) ;
\draw [shift={(304,210)}, rotate = 270] [fill={rgb, 255:red, 0; green, 0; blue, 0 }  ][line width=0.08]  [draw opacity=0] (8.93,-4.29) -- (0,0) -- (8.93,4.29) -- cycle    ;
\draw    (286,134) .. controls (271.3,94.8) and (319.99,90.17) .. (306.88,133.3) ;
\draw [shift={(306,136)}, rotate = 289.18] [fill={rgb, 255:red, 0; green, 0; blue, 0 }  ][line width=0.08]  [draw opacity=0] (8.93,-4.29) -- (0,0) -- (8.93,4.29) -- cycle    ;
\draw    (275.52,129.68) .. controls (245.24,58.05) and (350.46,60.13) .. (315,134) ;
\draw [shift={(277,133)}, rotate = 244.98] [fill={rgb, 255:red, 0; green, 0; blue, 0 }  ][line width=0.08]  [draw opacity=0] (8.93,-4.29) -- (0,0) -- (8.93,4.29) -- cycle    ;
\draw (265,178) node [anchor=north west][inner sep=0.75pt]   [align=left] {$ \displaystyle I$};
\draw (288,216) node [anchor=north west][inner sep=0.75pt]   [align=left] {$\displaystyle N$};
\draw (312,175) node [anchor=north west][inner sep=0.75pt]   [align=left] {$ \displaystyle J$};
\draw (290,145) node [anchor=north west][inner sep=0.75pt]   [align=left] {$\displaystyle n$};
\draw (290,83) node [anchor=north west][inner sep=0.75pt]   [align=left] {$\displaystyle B_{1}$};
\draw (290,51) node [anchor=north west][inner sep=0.75pt]   [align=left] {$\displaystyle B_{2}$};
\end{tikzpicture}
\label{ADHM-quiver}
\end{align}
ADHM数据满足$[B_1,B_2]+IJ=0$及$[B_1,B_1^\dagger]+[B_2,B_2^\dagger]+II^\dagger-J^\dagger J=\mu\ {\rm id}$的限制条件，而瞬子配分函数的表达式可以从计算（$\Omega$背景旋转作用叠加规范变换的）固定点附近的切空间上U($n$)规范变换作用得到\cite{Losev:1997tp,Nekrasov:2002qd}。对于涡旋解，也存在类似的构建法，更直接地，我们可以从瞬子解的ADHM数据中提取关于一个特定U(1)电荷中性的部分数据构造涡旋解\cite{Hanany:2003hp}，对于\eqref{ADHM-quiver}中的箭图，涡旋解仅保留$B_1$与$I$，限制条件也简化为$[B_1,B_1^\dagger]+II^\dagger=\mu\ {\rm id}$。

当给定一个箭图，我们就能通过考虑箭图中所有节点（node）之间由箭头指定的映射，模去它们之间的限制条件与等价关系，构造对应的中岛箭图代数簇（Nakajima quiver variety）\cite{quiver-variety}（更多信息可参考讲义\cite{Ginzburg}）。这类几何结构给出的便是包含瞬子及涡旋模空间的一类几何体的系统性构建方法。然而令人惊奇的是，以Maulik和Okounkov（MO）为首的数学家（受物理上Alday-Gaiotto-Tachikawa猜想\cite{Alday:2009aq,Wyllard:2009hg}的启发）率先发现了一个有趣的事实：中岛代数簇的等变上同调（equivariant cohomology）给出了箭图所对应的杨代数/量子群的表示\cite{Maulik-Okounkov,Schiffmann:2012tbu}。一个与本文主题息息相关的例子是\eqref{linear-quiver}中的线性箭图，
\begin{align}
\begin{tikzpicture}[x=0.75pt,y=0.75pt,yscale=-1,xscale=1]
\draw   (179,110.5) .. controls (179,105.25) and (183.25,101) .. (188.5,101) .. controls (193.75,101) and (198,105.25) .. (198,110.5) .. controls (198,115.75) and (193.75,120) .. (188.5,120) .. controls (183.25,120) and (179,115.75) .. (179,110.5) -- cycle ;
\draw   (221,110.5) .. controls (221,105.25) and (225.25,101) .. (230.5,101) .. controls (235.75,101) and (240,105.25) .. (240,110.5) .. controls (240,115.75) and (235.75,120) .. (230.5,120) .. controls (225.25,120) and (221,115.75) .. (221,110.5) -- cycle ;
\draw   (262,110.5) .. controls (262,105.25) and (266.25,101) .. (271.5,101) .. controls (276.75,101) and (281,105.25) .. (281,110.5) .. controls (281,115.75) and (276.75,120) .. (271.5,120) .. controls (266.25,120) and (262,115.75) .. (262,110.5) -- cycle ;
\draw   (350,109.5) .. controls (350,104.25) and (354.25,100) .. (359.5,100) .. controls (364.75,100) and (369,104.25) .. (369,109.5) .. controls (369,114.75) and (364.75,119) .. (359.5,119) .. controls (354.25,119) and (350,114.75) .. (350,109.5) -- cycle ;
\draw   (181,170) -- (199,170) -- (199,188) -- (181,188) -- cycle ;
\draw   (221,171) -- (239,171) -- (239,189) -- (221,189) -- cycle ;
\draw   (261,171) -- (279,171) -- (279,189) -- (261,189) -- cycle ;
\draw   (351,171) -- (369,171) -- (369,189) -- (351,189) -- cycle ;
\draw    (186.5,166) -- (186.5,125) ;
\draw [shift={(186.5,122)}, rotate = 90] [fill={rgb, 255:red, 0; green, 0; blue, 0 }  ][line width=0.08]  [draw opacity=0] (8.93,-4.29) -- (0,0) -- (8.93,4.29) -- cycle    ;
\draw    (194,123) -- (194,166) ;
\draw [shift={(194,169)}, rotate = 270] [fill={rgb, 255:red, 0; green, 0; blue, 0 }  ][line width=0.08]  [draw opacity=0] (8.93,-4.29) -- (0,0) -- (8.93,4.29) -- cycle    ;
\draw    (234,123) -- (234,166) ;
\draw [shift={(234,169)}, rotate = 270] [fill={rgb, 255:red, 0; green, 0; blue, 0 }  ][line width=0.08]  [draw opacity=0] (8.93,-4.29) -- (0,0) -- (8.93,4.29) -- cycle    ;
\draw    (274.5,123) -- (274.5,166) ;
\draw [shift={(274.5,169)}, rotate = 270] [fill={rgb, 255:red, 0; green, 0; blue, 0 }  ][line width=0.08]  [draw opacity=0] (8.93,-4.29) -- (0,0) -- (8.93,4.29) -- cycle    ;
\draw    (363.5,122) -- (363.5,165) ;
\draw [shift={(363.5,168)}, rotate = 270] [fill={rgb, 255:red, 0; green, 0; blue, 0 }  ][line width=0.08]  [draw opacity=0] (8.93,-4.29) -- (0,0) -- (8.93,4.29) -- cycle    ;
\draw    (225.5,165) -- (225.5,124) ;
\draw [shift={(225.5,121)}, rotate = 90] [fill={rgb, 255:red, 0; green, 0; blue, 0 }  ][line width=0.08]  [draw opacity=0] (8.93,-4.29) -- (0,0) -- (8.93,4.29) -- cycle    ;
\draw    (265.5,164) -- (265.5,123) ;
\draw [shift={(265.5,120)}, rotate = 90] [fill={rgb, 255:red, 0; green, 0; blue, 0 }  ][line width=0.08]  [draw opacity=0] (8.93,-4.29) -- (0,0) -- (8.93,4.29) -- cycle    ;
\draw    (354.5,164) -- (354.5,123) ;
\draw [shift={(354.5,120)}, rotate = 90] [fill={rgb, 255:red, 0; green, 0; blue, 0 }  ][line width=0.08]  [draw opacity=0] (8.93,-4.29) -- (0,0) -- (8.93,4.29) -- cycle    ;
\draw    (198,106) -- (217.5,106) ;
\draw [shift={(220.5,106)}, rotate = 180] [fill={rgb, 255:red, 0; green, 0; blue, 0 }  ][line width=0.08]  [draw opacity=0] (8.93,-4.29) -- (0,0) -- (8.93,4.29) -- cycle    ;
\draw    (240,105) -- (259.5,105) ;
\draw [shift={(262.5,105)}, rotate = 180] [fill={rgb, 255:red, 0; green, 0; blue, 0 }  ][line width=0.08]  [draw opacity=0] (8.93,-4.29) -- (0,0) -- (8.93,4.29) -- cycle    ;
\draw    (284,105) -- (303.5,105) ;
\draw [shift={(306.5,105)}, rotate = 180] [fill={rgb, 255:red, 0; green, 0; blue, 0 }  ][line width=0.08]  [draw opacity=0] (8.93,-4.29) -- (0,0) -- (8.93,4.29) -- cycle    ;
\draw    (327,105) -- (346.5,105) ;
\draw [shift={(349.5,105)}, rotate = 180] [fill={rgb, 255:red, 0; green, 0; blue, 0 }  ][line width=0.08]  [draw opacity=0] (8.93,-4.29) -- (0,0) -- (8.93,4.29) -- cycle    ;
\draw    (222,114) -- (200,114) ;
\draw [shift={(197,114)}, rotate = 360] [fill={rgb, 255:red, 0; green, 0; blue, 0 }  ][line width=0.08]  [draw opacity=0] (8.93,-4.29) -- (0,0) -- (8.93,4.29) -- cycle    ;
\draw    (307,115) -- (285,115) ;
\draw [shift={(282,115)}, rotate = 360] [fill={rgb, 255:red, 0; green, 0; blue, 0 }  ][line width=0.08]  [draw opacity=0] (8.93,-4.29) -- (0,0) -- (8.93,4.29) -- cycle    ;
\draw    (263,114) -- (241,114) ;
\draw [shift={(238,114)}, rotate = 360] [fill={rgb, 255:red, 0; green, 0; blue, 0 }  ][line width=0.08]  [draw opacity=0] (8.93,-4.29) -- (0,0) -- (8.93,4.29) -- cycle    ;
\draw    (349,116) -- (327,116) ;
\draw [shift={(324,116)}, rotate = 360] [fill={rgb, 255:red, 0; green, 0; blue, 0 }  ][line width=0.08]  [draw opacity=0] (8.93,-4.29) -- (0,0) -- (8.93,4.29) -- cycle    ;
\draw (304,100) node [anchor=north west][inner sep=0.75pt]   [align=left] {$\displaystyle \dotsc $};
\draw (181,72) node [anchor=north west][inner sep=0.75pt]   [align=left] {$\displaystyle v_{1}$};
\draw (224,73) node [anchor=north west][inner sep=0.75pt]   [align=left] {$\displaystyle v_{2}$};
\draw (264,73) node [anchor=north west][inner sep=0.75pt]   [align=left] {$\displaystyle v_{3}$};
\draw (350,74) node [anchor=north west][inner sep=0.75pt]   [align=left] {$\displaystyle v_{r}$};
\draw (180,199) node [anchor=north west][inner sep=0.75pt]   [align=left] {$\displaystyle w_{1}$};
\draw (219,200) node [anchor=north west][inner sep=0.75pt]   [align=left] {$\displaystyle w_{2}$};
\draw (261,200) node [anchor=north west][inner sep=0.75pt]   [align=left] {$\displaystyle w_{3}$};
\draw (349,198) node [anchor=north west][inner sep=0.75pt]   [align=left] {$\displaystyle w_{r}$};
\end{tikzpicture}
\label{linear-quiver}
\end{align}
它与箭图规范场论（quiver gauge theory）中的涡旋解对应，而其等变上同调同构于
\begin{equation}
    H_T^\bullet(X)\simeq \lt(\mathbb{C}^n\rt)^{\otimes w_1}\otimes \lt(\wedge^2\mathbb{C}^n\rt)^{\otimes w_2}\otimes \dots \otimes \lt(\wedge^r\mathbb{C}^n\rt)^{\otimes w_r}.
\end{equation}
这等价于考虑杨代数/量子群$U_q(\widehat{\mathfrak{sl}}_{r+1})$各类表示的直积，正好与贝特/规范对偶中在自旋链的不同格点处插入$\mathfrak{su}(r+1)$内部自由度的不同表示相对应。于是乎，我们看到物理上偶然发现的看似毫无关联的超对称规范场论与量子可积模型之间的对应关系，可以通过数学上的几何表示论方法予以系统性的阐释和研究。这其中图像最为清晰的系统之一便是\eqref{ADHM-quiver}所对应的四维理论瞬子模空间上的量子代数结构。该箭图所对应的量子代数\footnote{这里不得不承认的是，箭图与代数之间的对应是一个很难通过物理角度解释的数学结论。在代数表示论的框架中，量子代数其实是通过箭图代数簇上定义的R矩阵分析得到的\cite{Maulik-Okounkov,Okounkov:2016sya}。从物理的角度出发，也可以利用BPS箭图提取量子代数\cite{Li:2020rij}。}通常被称为$\widehat{\mathfrak{gl}}_1$的仿射杨代数（affine Yangian）/量子环面代数（quantum toroidal algebra）或Ding-Iohara-Miki代数\cite{Ding:1996mq,miki2007q}，感兴趣的读者可以参考综述论文\cite{Matsuo:2023lky}。这个代数最有趣的性质是它是一个准三角（quasi-triangular）Hopf代数，Hopf代数意味着代数（记为$H$）中存在一个余积（coproduct）结构
：$\Delta:\ H\to H\otimes H$，以及余积与置换操作的合成，$\Delta^{op}={\cal P}\circ \Delta$。准三角的定义为，存在$R\in{\rm End}(H\otimes H)$，满足以下性质：
\begin{align}
    &R\Delta(g)=\Delta^{op}(g)R,\quad ^\forall g\in H,\label{R-coprod}\\
    &({\rm id}\otimes \Delta)R=R_{13}R_{12},\quad (\Delta\otimes {\rm id})R=R_{13}R_{23}
\end{align}
这个$(H\otimes H)$代数中的元$R$给出一个YBE的解，这里我们给一个简要证明：首先将$R$分解为$R=\sum_i a_i\otimes b_i$，之后便易证，
\begin{equation}
    \begin{aligned}
R_{12} R_{13} R_{23} & =R_{12}(\Delta \otimes \textrm{id}) R  =\sum_i(R \Delta(a_i)) \otimes b_i   =\sum(\Delta^{op}(a_i) R) \otimes b_i  =({\cal P} \otimes \textrm{id}) \cdot(\Delta \otimes \textrm{id}) R \cdot R_{12}\\
& =({\cal P} \otimes \textrm{id})(R_{13} R_{23})R_{12}  =R_{23} R_{13} R_{12}.
\end{aligned}
\end{equation}
上述$R$被称为普遍R-矩阵（universal R-matrix），并且可以通过求解\eqref{R-coprod}确定$R$的具体表达式。$\widehat{\mathfrak{gl}}_1$的量子环面代数中的普遍R-矩阵在杨代数极限下与MO在\cite{Maulik-Okounkov}中通过几何表示论发现的R-矩阵表达式一致\cite{Fukuda:2017qki}，并且可以通过二维刘维尔共性场论中的能动量张量定义：
\begin{align}
    &R(u)T(Q)=T(-Q)R(u),\quad T(Q)=\frac{1}{2}\partial\varphi^-\partial\varphi^-+Q\partial^2\varphi^-,\\
    &R(u)|\eta_1\rangle\otimes |\eta_2\rangle=|\eta_1\rangle\otimes |\eta_2\rangle,\quad u=\frac{1}{\sqrt{2}}(\eta_1-\eta_2),\quad a_0|\eta\rangle=\eta |\eta\rangle,
\end{align}
其中$\varphi(z)=p_0+a_0\log z-\sum_{n\in\mathbb{Z}}\frac{a_n}{n}z^{-n}$为自由玻色子，而$\varphi^-:=\frac{1}{\sqrt{2}}(\varphi^{(1)}-\varphi^{(2)})$为R-矩阵作用的两个Fock空间上的自由玻色子之差。当考虑上述R-矩阵对应的定义在一个Fock量子空间上的单格点量子可积模型时，其哈密顿量为
\begin{equation}
    {\cal H}=\sum_{n,m>0}(a_{-n}a_{-m}a_{m+n}+a_{-n-m}a_na_m)+\sqrt{2}Q\sum_{n>0}na_{-n}a_n,
\end{equation}
而这正是粒子数无穷极限下的Calogero-Sutherland模型的哈密顿量表达式\cite{Awata:1994xd}。对应的多格点版本量子可积模型中会进一步引入格点之间不同玻色子之间的相互作用（感兴趣的读者可以参考\cite{Maulik-Okounkov,Zhu:2015nha}）。

最后让我们梳理一下如何从四维超对称场论出发建立与自旋链模型之间的贝特/规范对应。这其中分为两步，第一步取$\vec{a}=-\vec{m}-\epsilon_1\vec{n}$将瞬子配分函数约化为二维理论的涡旋配分函数；第二步进一步取$\epsilon\equiv\epsilon_2\to 0$极限回到二维理论在平坦时空背景下的低能有效物理，其中可以提取出自旋链模型的信息。定性地，这两步操作在量子代数中同样能还原出自旋链。在第一步中，标记瞬子解的杨图被限制到了有限行，这对应着Calogero-Sutherland模型中将粒子数限制到有限。第二步操作对应于模型中取$\beta\equiv -\frac{\epsilon_1}{\epsilon_2}\to\infty$的强耦合极限，常被称为冻结极限（freezing limit）\cite{Bernard:1993va}，可得到一种可积的长程自旋格点模型——Haldane-Shastry模型\cite{Haldane:1987gg,SriramShastry:1987wdh}。后者又通过Inozemtsev椭圆模型的不同极限与自旋链模型建立对应\cite{Inozemtsev}，这样我们成功从模空间上的量子可积性角度为贝特/规范对应关系提供了一种大致解释。

\section{总结与展望}

本文回顾了贝特/规范对应关系中场论的真空方程与可积模型中BAE之间的等价关系，并通过涡旋配分函数的角度对这套偶然发现的对应关系给予了一定的解释，将其中的量子可积性最终归结至四维$\cN=2$理论中普遍存在的量子代数结构及其背后的可积性在特殊极限下的展现。然而由于篇幅限制，本文在贝特/规范对应的文脉中也只能阐述部分侧面，在文章的最后让我们简要提及一下文献中关联密切的部分工作。在Nekrasov和Shatashvili的最初工作中便提出了贝特/规范对应可推广至SO或Sp规范理论与开边界自旋链，与对角开边界模型之间具体的对应词典在\cite{Kimura:2020bed}中被给出，随后在\cite{Ding:2023auy,Ding:2023lsk,Ding:2023nkv,Wang:2024zcr}等工作被推广到了任意规范群及非对角开边界条件等更一般的情况（对应的贝特拟设解法最早由\cite{Cao:2013nza}提出）。类似的与自旋链之间的对应关系还有通过膜结构（brane construction）建立的四维$\cN=1$理论的Gauge/YBE对应\cite{Yamazaki:2013nra,Yagi:2015lha}、通过四维Chern-Simons理论提取出的自旋链信息\cite{Costello:2017dso,Costello:2018txb}。对于三维Chern-Simons理论，同样可建立贝特/规范对应，但对应的可积模型也定义在Fock空间上\cite{Korff-Stroppel,Okuda:2013fea}，较为复杂却有趣。在量子代数方面，近年来最大的进展之一便是发现超对称规范场论的BPS箭图与代数结构存在一对一对应关系\cite{Li:2020rij,Galakhov:2021xum,Galakhov:2022uyu}，但似乎并非在所有的类环面代数结构中均存在R-矩阵及量子可积结构，还有待进一步挖掘。

以贝特/规范对应为代表的超对称规范场论中的量子可积性仍有许多问题有待解决。量子可积性在量子场论中的物理来源并不完全清晰。SW曲线的来源通过场论的膜构造、M理论及F理论紧化在Calabi-Yau流形上的几何工程（geometric engineering）得到解释，本文中也提到量子可积性的出现可以视作一种SW曲线的量子化甚至双重量子化过程。遗憾的是，目前并不清楚是否所有具有SW曲线描述的量子场论均可自洽地进行量子化，比如一个有名的事实是量子环面代数不能直接写出BCD型（参见\cite{Awata:2017lqa}中的讨论），尽管代数表示论的研究中似乎能抽象地定义相关的可积模型\cite{Nakajima:2019olw}。另一方面，与AdS/CFT对偶中运用大量自旋链计算技巧求解$\cN=4$物理信息不同，$\cN=2$理论的可积性研究中多数工作还聚焦于如何运用局域化手法计算得到的场论结果还原自旋链中的已知信息。自旋链模型的研究已有近百年历史，可以运用的成熟研究算法颇多（尤其是数值手法，例如最近开发的方法： \cite{Marboe:2016yyn,Hou:2023ndn}），如何将它们巧妙地用于场论中的计算，将是接下来研究的一大课题。希望本文的撰写能吸引更多有生力量加入到这个有趣但略为小众的研究领域中。

\paragraph{致谢}
RZ非常感谢A. Smirnov对相关话题上耐心的答疑释惑，以及G. Cotti, G. Bonelli与A. Tanzini组织的训练学校CTIS24，本文的撰写工作部分是在相关学术活动中推进的，并得到了极大的启发。本工作由国家自然科学基金委青年基金项目（No. 12105198）支持。另外，作者想特别感谢《中国科学: 物理学 力学 天文学》的审稿人，其细致入微的建议大大增加了本文的专业性，也让作者学习到了很多难以直接从文献中精准抓取的经典文献中的知识。

{\scriptsize
\bibliography{Bethe.bib}
}

\end{CJK*}

\end{document}